\documentclass[floats,aps,prl,eqnum,showpacs,twocolumn]{revtex4}

\usepackage{subfigure}
\usepackage[dvips]{color,graphicx}
\usepackage{amsfonts,amssymb,theorem}
\newcommand{\ma}[1]{\mbox{$\mathcal{#1}$}}

{\theorembodyfont{\upshape}
}
{\theorembodyfont{\upshape}
}
{\theorembodyfont{\upshape}
}
{\theorembodyfont{\upshape}
}
{\theorembodyfont{\upshape}
}
{\theorembodyfont{\upshape}
}

\newcommand{\dalm}{\kern1pt\vbox{\hrule height 0.9pt\hbox{\vrule width
0.9pt\hskip 2.5pt\vbox{\vskip 5.5pt}\hskip 3pt\vrule width 0.3pt}\hrule height
0.3pt}\kern1pt}

\begin{document}

\title{
Global structure and physical interpretation of the Fonarev solution for a scalar field with exponential potential}

\author{Hideki Maeda$^{1,2}$}
\email{hideki@cecs.cl}


\address{ 
$^{1}$Centro de Estudios Cient\'{\i}ficos (CECS), Arturo Prat 514, Valdivia, Chile\\
$^{2}$Department of Physics, International Christian University, 3-10-2 Osawa, Mitaka-shi, Tokyo 181-8585, Japan
}

\date{\today}

\begin{abstract} 
We discuss the physical interpretation of a dynamical and inhomogeneous spherically symmetric solution obtained by Fonarev for a scalar field with an exponential potential.
There is a single parameter $w$ in the solution which can be set to $\pm1$ if it is non-zero, in addition to the steepness parameter $\lambda$ in the potential.
The spacetime is conformally static and asymptotically flat Friedmann-Robertson-Walker spacetime.
The solution reduces to the Friedmann-Robertson-Walker solution for $w=0$. 
There are two curvature singularities, of which one is a timelike central singularity and the other is a big-bang or big-crunch type singularity.
Depending on the parameters, the spacetime can possess a future outer trapping horizon in the collapsing case.
Then the solution represents a dynamical black hole in the sense of Hayward although there is a locally naked singularity at the center and no black-hole event horizon.
This demonstrates a weak point of the local definition of a black hole in terms of a trapping horizon.
\end{abstract}

\maketitle

Einstein equations are so complicated simultaneous nonlinear partial differential equations that it is hopeless to obtain general solutions.  
Thus, spacetime symmetries are often assumed to make the system more tractable.
Although such spacetimes with high symmetries are idealized ones, they have occupied an important position in the history of gravitation research as touchstones to know the essential physics~\cite{exactsolutions,krasinski}.

For example, the Friedmann-Robertson-Walker (FRW) cosmological model plays a central role in modern cosmology as the zeroth-order approximation of the present universe, on which the behavior of the density perturbations has been particularly investigated to clarify the origin of the large-scale structure of the universe or to determine the cosmological parameters from the observations of the cosmic microwave background~\cite{peebles,linde,liddle}.
From the analyses of the stationary and asymptotically flat black-hole spacetime such as the Schwarzschild or Kerr solution, tremendous results have been derived, among which the most remarkable one is the black-hole thermodynamics~\cite{he1973,wald1983}. 

On the other hand, there are many open problems on the dynamical aspects of Einstein equations such as dynamical black holes or their formations.
In homogeneous or stationary spacetimes, Einstein equations reduce to a set of ordinary differential equations, which is comparatively easy to handle, however, the formation or the growth of a black hole is essentially a dynamical and inhomogeneous process, where we have to struggle with a set of partial differential equations.
For this reason, numerical methods have often been used to study such systems.
Nevertheless, the few dynamical and inhomogeneous exact solutions have been found.
Such precious solutions should be intensively investigated to complement the numerical works.

Scalar fields are fundamental fields which naturally exist in a variety of theories.
In spherically symmetric spacetimes, the system with the simplest massless scalar field has been fully investigated.
There are two important dynamical and inhomogeneous exact solutions in this case.
The first one was obtained by Roberts~\cite{roberts1989}, and subsequently the other one was obtained by Husain, Martinez and N{\'u}{\~n}ez~\cite{hmn1994}.
The dynamical aspects of the system with a massless scalar field have been investigated in the numerical studies of the gravitational collapse, especially in the context of critical phenomena pioneered by Choptuik~\cite{choptuik1993} (see~\cite{gundlach} for the review).
The analytic proof of the cosmic censorship hypothesis by Christodoulou is a significant milestone~\cite{christodoulou}.
Although these two exact solutions are not directly related to these results, the potential importance of them would go without saying.

The case with potentials is a natural generalization of the massless case.
In various theories, scalar fields have their specific potentials.
Among them, an exponential potential arises naturally in supergravity~\cite{townsend2001} or theories obtained through dimensional reduction to effective four-dimensional theories~\cite{gl2000,eg2003}.
Indeed, the existence of exact solutions in such systems must be useful as a touchstone for the future research.
The generalized Husain-Martinez-N{\'u}{\~n}ez (HMN) solution in the presence of an exponential potential was obtained by Fonarev~\cite{fonarev1995}.
However, the properties of the solution have not been studied in details and the physical interpretation of the solution is still not clear.
In this letter, we discuss the physical interpretation of the Fonarev solution.
We investigate the properties of the trapping horizon and show the global structure of the solution.
We adopt the units such that $c=G=1$.

We begin with the action which describes the system with a scalar field with an exponential potential: 
\begin{equation}
S=\int d^4x\sqrt{-g}\left[\frac{1}{8\pi}{\ma R}-\left(\frac12 \phi_{,\mu}\phi^{,\mu}+V(\phi)\right)\right],
\label{action}
\end{equation}
where $V$ is the potential of a scalar field $\phi$ given by 
\begin{eqnarray}
V(\phi) := V_{0}e^{-\sqrt{8\pi} \lambda\phi},
\end{eqnarray}
where $V_0$ and $\lambda$ are real constants. 
A scalar field with an exponential potential has been particularly studied in spatially homogeneous cosmology~\cite{expcosmology}.
If $V_{0}=0$, then $\phi$ is massless and $\lambda$ is meaningless. 
We assume that $V_0$ is non-negative.

Then the energy-momentum tensor for a scalar field is given by 
\begin{equation}
T_{\mu\nu}=\left(\phi_{,\mu}\phi_{,\nu}-\frac{1}{2}g_{\mu\nu}\phi_{,\rho}\phi^{,\rho}\right)-g_{ab}V.
\label{eq:stress-energy_tensor_of_scalar_field}
\end{equation}
The Einstein equation is 
\begin{equation}
{\ma R}_{\mu\nu}=8\pi\left(\phi_{,\mu}\phi_{,\nu}+g_{\mu\nu}V\right),
\end{equation}
while equations of motion for $\phi$ is given by 
\begin{eqnarray}
\dalm\phi=\frac{\partial V}{\partial \phi}.\label{eq:Klein-Gordon1}
\end{eqnarray}

The spherically symmetric solution obtained by Fonarev~\cite{fonarev1995} is 
\begin{eqnarray}
ds^2&=&a(\eta)^2\left[-f(r)^2d\eta^2+\frac{dr^2}{f(r)^2}+S(r)^2d\Omega^2\right], \label{sol} \\
\phi &=& \frac{1}{4\sqrt{\pi}}\ln\biggl[d\biggl(1-\frac{2w}{r}\biggl)^{\sqrt{2/(\lambda^2+2)}}a^{\sqrt{2}\lambda}\biggl],\label{sol2}
\end{eqnarray}
where
\begin{eqnarray}
f(r) &:=& \left(1-\frac{2w}{r}\right)^{\alpha/2},\\
S(r)^2 &:=& r^2\left(1-\frac{2w}{r}\right)^{1-\alpha},\\
\alpha&:=&\frac{\lambda}{\sqrt{\lambda^2+2}},\\
a(\eta)&:=&\biggl|\frac{\eta}{\eta_0}\biggl|^{2/(\lambda^2-2)},\\
d&:=&\biggl[\frac{2(6-\lambda^2)}{8\pi V_0\eta_0^2(\lambda^2-2)^2}\biggl]^{-\sqrt{2}/\lambda}.
\end{eqnarray}
Here $w$ and $\eta_0$ are constants.
When we set $w=0$, the solution reduces to the FRW solution.
Hereafter we only consider the case with $w\ne 0$.

The potential form is then given by
\begin{eqnarray}
V_0e^{-\sqrt{8\pi}\lambda\phi}&=&\frac{6-\lambda^2}{4\pi \eta_0^2(\lambda^2-2)^2} \nonumber \\
&&\times\biggl(1-\frac{2w}{r}\biggl)^{-\lambda/\sqrt{\lambda^2+2}}a^{-\lambda^2},
\end{eqnarray}
so the steepness parameter $\lambda$ must satisfy the relation $0<\lambda^2 \le 6$ and $\lambda^2 \ne 2$ for non-negative potential, which corresponds to $0< \alpha^2 \le 3/4$ with $\alpha^2 \ne 1/2$.
This solution is asymptotically FRW solution for $r \to \infty$.

For $0<\lambda^2<2$ (corresponding to $0<\alpha^2<1/2$), the asymptotic FRW solution is accelerated, while for $2<\lambda^2 \le 6$ (corresponding to $1/2<\alpha^2 \le 3/4$), it is decelerated.
When we set $\lambda^2=6$, this solution reduces to the HMN solution for a massless scalar field, in which $d$ is an arbitrary constant and meaningless~\cite{hmn1994}.

Because this solution is invariant for $r \to -r$ with $w \to -w$, we consider only the region of $r\ge 0$.
For $w \ne 0$, there are central curvature singularities at $r=0$ and $2w$, which are both timelike.
Thus, the domain of definition for $r$ is $2w < r<+\infty$ for $w > 0$, while it is $0 < r<+\infty$ for $w < 0$.
Then, we can set $w=+(-)1$ for positive (negative) $w$ without loss of generality by the coordinate transformations ${\tilde \eta}:=\eta/|w|$ and ${\tilde r}:=r/|w|]$ and the redefinition of the constant $\eta_0$.

In addition to them, $\eta \to \pm\infty$ is a null curvature singularity for $0<\lambda^2<2$, in which case $\eta=+(-)0$ corresponds to past (future) infinity, while $\eta=0$ is a spacelike curvature singularity for $2<\lambda^2 \le 6$, in which case $\eta=+(-)\infty $ corresponds to future (past) infinity.
These are the big-bang or big-crunch type singularities. 
The domain of definition for $\eta$ is thus $-\infty< \eta<0$ and $0 < \eta<+\infty$.

The physical (areal) radius $R$ is given by $R:=aS$.
$R$ is a monotonically increasing function of $r$ for $2w < r<+\infty$ and $r \to \infty$ corresponds to the spacelike infinity.
For $0<\lambda^2<2$, $R$ is a monotonically increasing (decreasing) function of $\eta$ for $\eta<(>)0$, i.e., the spacetime is expanding (collapsing).
For $2<\lambda^2 \le 6$, on the other hand, it is a monotonically increasing (decreasing) function of $\eta$ for $\eta>(<)0$.

Here let us consider whether the scalar field has a non-trivial configuration or not.
We should be careful that there is no natural time-slicing in general spherically symmetric spacetimes, so the derivative of the scalar field must have a spacelike portion to have the non-trivial configuration in a correct sense.
For the Fonarev solution, we have
\begin{eqnarray}
16\pi a^2\phi_{,\mu}\phi^{,\mu}&=& -\frac{8\lambda^2}{(\lambda^2-2)^2}\frac{1}{\eta^2}\biggl(1-\frac{2w}{r}\biggl)^{-\alpha} \nonumber \\
&&+\frac{2}{\lambda^2+2}\frac{4w^2}{r^4}\biggl(1-\frac{2w}{r}\biggl)^{\alpha-2},
\end{eqnarray}
and there is indeed a region with $\phi_{,\mu}\phi^{,\mu}>0$.

The Misner-Sharp mass $m$ is given by
\begin{eqnarray}
m=\frac{R}{2}(1-R_{,\mu}R^{,\mu}),
\end{eqnarray}
where we have
\begin{eqnarray}
R_{,\mu}R^{,\mu}&=&-\left(\frac{a_{,\eta}}{a}\right)^2r^2\biggl(1-\frac{2w}{r}\biggl)^{1-2\alpha} \nonumber \\
&&+\biggl(1-\frac{(1+\alpha)w}{r}\biggl)^2\biggl(1-\frac{2w}{r}\biggl)^{-1}.
\end{eqnarray}
It is noted that 
\begin{eqnarray}
\lim_{r \to 2w}m=-\infty
\end{eqnarray}
and 
\begin{eqnarray}
\lim_{r \to 0}m=-\infty
\end{eqnarray}
for $0< \alpha^2 \le 3/4$ with $\alpha^2 \ne 1/2$.

A trapping horizon is obtained by $R=2m$, or equivalently $R_{,\mu}R^{,\mu}=0$, which is given by
\begin{eqnarray}
0&=&\frac{1}{\eta}\pm\biggl|\frac{\lambda^2-2}{2}\biggl|\frac{1}{r}\left(1-\frac{2w}{r}\right)^{\alpha-1} \nonumber \\
&&~~~~~~~~~~~\times\biggl(1-\frac{(1+\alpha)w}{r}\biggl),\label{T-horizon} \\
&=:&h(\eta,r).
\end{eqnarray}
It is noted that outgoing (ingoing) null geodesics are trapped on the trapped regions in the collapsing (expanding) regions.
The normal vector $l^\mu$ to the surface $h=0$ is given by $l^\mu=g^{\mu\nu}h_{,\nu}$.
We obtain 
\begin{eqnarray}
l^\mu l_\mu&=&-\frac{(1-2w/r)^{-\alpha} [\alpha r-(1+\alpha)w]^2g(r)}{a^2(2\alpha^2-1)^2[r-(1+\alpha)w]^4},\\
g(r)&:=&(3\alpha^2-2)r^2 \nonumber \\
&&-2(1+\alpha)(4\alpha^2-\alpha-2)wr \\
&&+(4\alpha^2-3)(1+\alpha)^2w^2
\end{eqnarray}
for both signs in Eq.~(\ref{T-horizon}).
The function $g$ is negative for $0<\alpha^2 < 1/2$, which implies that the trapping horizon is timelike.
The case of $1/2 < \alpha^2 \le 3/4$ is rather complicated.
We obtain
\begin{eqnarray}
g(2w)=(4\alpha^2-3)(1-\alpha)^2w^2,
\end{eqnarray}
which is non-positive for $-\sqrt{3}/2 \le \alpha \le \sqrt{3}/2$.
The solution of $g=0$ exists only for $\alpha^2 \ge 1/2$.
For $\alpha = \pm\sqrt{2/3}$, we have $g=\pm2\sqrt{6}wr/9-(5\pm2\sqrt{6})w^2/9$.
Thus, for $\alpha = -\sqrt{2/3}$, the trapping horizon is timelike for $r>2w$, while for $\alpha = \sqrt{2/3}$, the trapping horizon is timelike (spacelike) for $r<(>)r_0$, where $r_0$ is defined by $r_{0}:=\sqrt{6}(5+2\sqrt{6})w/12$.

Next let us consider the case with $\alpha \ne \pm\sqrt{2/3}$.
The solutions of $g=0$ are then given as $r=r_{\rm ex(+)}, r_{\rm ex(-)}$ defined by
\begin{widetext}
\begin{eqnarray}
r_{\rm ex(\pm)}:=\frac{(1+\alpha)w[8\alpha^2-2\alpha-4\pm 2(1-\alpha)\sqrt{2(2\alpha^2-1)}]}{2(3\alpha^2-2)}.
\end{eqnarray}
\end{widetext}
We can show that $r_{\rm ex(+)}\ge 2w$ holds for $-\sqrt{3}/2 \le \alpha < -\sqrt{2/3}$, $\sqrt{1/2} < \alpha <\sqrt{2/3}$, and $\sqrt{2/3} < \alpha \le \sqrt{3}/2$ with equality holding for $\alpha=\pm\sqrt{3}/2$, while $r_{\rm ex(+)}< 2w$ holds for $-\sqrt{2/3} < \alpha < -\sqrt{1/2}$.
We can also show that $r_{\rm ex(-)}> 2w$ holds for $\sqrt{1/2} < \alpha < \sqrt{2/3}$, while $r_{\rm ex(-)}< 2w$ holds for $-\sqrt{3}/2 \le \alpha < -\sqrt{2/3}$, $-\sqrt{2/3}< \alpha < -\sqrt{1/2}$, and $\sqrt{2/3} < \alpha \le \sqrt{3}/2$.
Thus, it is concluded that the trapping horizon is timelike for $-\sqrt{2/3} < \alpha < -\sqrt{1/2}$ and spacelike for $\alpha=\pm\sqrt{3}/2$, while in other cases it has both timelike and spacelike portions.
For $2/3 < \alpha^2 < 3/4$, the trapping horizon is timelike for $2w < r < r_{\rm ex(+)}$ and spacelike for $r_{\rm ex(+)}< r$.
For $\sqrt{1/2} < \alpha < \sqrt{2/3}$, the trapping horizon is timelike for $2w < r < r_{\rm ex(-)}$ and $r_{\rm ex(+)} < r$, while it is spacelike for $r_{\rm ex(-)}< r < r_{\rm ex(+)}$.

We have now found that the properties of the trapping horizon are different from that in the HMN solution corresponding to $\alpha=\pm\sqrt{3}/2$, which is spacelike.
A local definition of a black hole in terms of a trapping horizon was given by Hayward~\cite{hayward1994}.
Then, a future outer trapping horizon is a black-hole horizon, which corresponds to a spacelike trapping horizon in the collapsing region.
On the other hand, a spacelike trapping horizon in the expanding region is a future inner trapping horizon corresponding to a white-hole or cosmological horizon.
Thus, there exists a black-hole trapping horizon for $-\sqrt{3}/2 \le \alpha < -\sqrt{2/3}$ (corresponding to $-\sqrt{6}\le \lambda<-\sqrt{2}$) and $\sqrt{1/2} < \alpha \le \sqrt{3}/2$ (corresponding to $\sqrt{2}< \lambda \le \sqrt{6}$).

\begin{figure}[htbp]
\begin{center}
\includegraphics[width=0.9\linewidth]{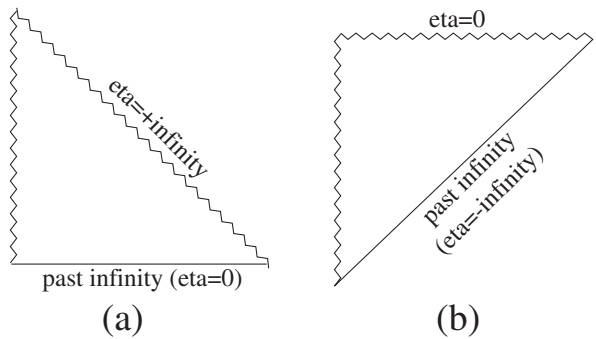}
\caption{\label{Penrose}
The Penrose diagrams for the Fonarev solution in the collapsing case with $w \ne 0$ and (a) $0<\lambda^2 <2$ and (b) $2<\lambda^2 \le 6$.
A zigzag line corresponds to a curvature singularity.
The central timelike singularity is at $r=2w$ and $r=0$ for $w>0$ and $w<0$, respectively.
For $w=0$, it is replaced by a regular center.
The diagrams in the expanding cases are obtained by setting the figures upside-down.
}
\end{center}
\end{figure}

In summary, the solution represents a dynamical black hole in the sense of Hayward with the non-trivial configuration of a scalar field, i.e., a scalar hair, in the collapsing case with $2< \lambda^2 \le 6$, where the asymptotic FRW solution is decelerating.
The Penrose diagrams of the Fonarev solution are given in Fig.~\ref{Penrose}.

In the static and asymptotically flat case, the scalar no-hair theorem is available, in which the scalar field must have a trivial configuration for a black-hole solution for the arbitrary positive semidefinite potential~\cite{heusler1992,bekenstein1995,sudarsky1995}.
This scalar no-hair theorem was extended into the asymptotically de~Sitter case in the presence of the convex potential~\cite{tmn1999a}. 
On the other hand, a counter example was numerically constructed in the asymptotically anti-de~Sitter case, which violates the null energy condition~\cite{tmn2001}.

If the scalar no-hair theorem also holds in general dynamical and inhomogeneous spherically symmetric spacetimes with certain energy conditions, a scalar field must have a trivial configuration for black-hole solutions.
The contraposition of this statement is that if a scalar field has a non-trivial configuration, the solution is not a black-hole solution.
Then, does the Fonarev solution suggest that the scalar no-hair theorem can be extended into general spherically symmetric spacetimes?
The answer is ``No'' in the sense of Hayward.
However, here we face with the subtlety of the definition of a black hole.

Although the solution represents a dynamical black hole in the sense of Hayward, the singularity inside that trapping horizon is the big-crunch type spacelike singularity, and the central singularity is timelike and not covered by trapped surfaces.
Furthermore, there is no black-hole event horizon in this spacetime since there is no future null infinity.
From these viewpoints, it might be proper to say that the Fonarev solution as well as the HMN solution represents not a black hole but a naked singularity.
Then, the answer to the above question turns to be ``Yes''.

These solutions may demonstrate a weak point of the local definition of a black hole in terms of a trapping horizon.
In order to rule out such solutions, we should require additional conditions in the definition by a trapping horizon or another local definition of a black hole.
Further investigations are needed to have much insight into this problem.

\section*{Acknowledgments}
The author would like to thank C.~Bunster, C.~Mart{\'{\i}}nez, R.~Troncoso, S.~Willison, and J.~Zanelli for useful comments. 
The author thanks Kjell Tangen for the crucial information of the paper by Fonarev.
This work was supported by the Grant Nos. 1071125 from FONDECYT (Chile) and the Grant-in-Aid for Scientific Research Fund of the Ministry of Education, Culture, Sports, Science and Technology, Japan (Young Scientists (B) 18740162).


\end{document}